\documentstyle[12pt]{article}
\begin{document}
\begin{center}
\LARGE
Scaling of the Random-Field Ising Model at Zero Temperature
 ~\\
 ~\\
 ~\\
\Large
Michael R. Swift,$^1$ Alan J. Bray, $^2$ Amos Maritan,$^1$
Marek Cieplak,$^{3}$ and Jayanth R. Banavar$^{4}$
 ~\\
 ~\\
\normalsize
{\it
$^{1}$Istituto Nazionale di Fisica della Materia, \\
International School for Advanced Studies, \\
I-34014 Grignano di
Trieste and sezione INFN di Trieste, Italy \\
~\\
$^2$ Department of Theoretical Physics, University of Manchester,
Manchester M13 9PL, UK, \\
~\\
$^3$ Institute of Physics, Polish Academy of Sciences, 02-668
Warsaw, Poland, \\
~\\
$^{4}$Department of Physics and Center for Materials Physics \\
The Pennsylvania State University \\
104 Davey Laboratory, University Park, PA  16802 USA}
\end{center}

\vskip 0.4in

\addtolength{\baselineskip}{\baselineskip}

\begin{center}
Abstract
\end{center}

The exact determination of ground states of small systems
is used in a scaling study of the random-field Ising model.
While three variants of the model are found to be in the same
universality class in 3 dimensions, the Gaussian and bimodal
models behave distinctly in 4 dimensions with the
latter apparently having a discontinuous jump in the magnetization.
A finite-size scaling analysis is presented for this transition. 

~

\newpage
The random-field Ising model (RFIM) has been a subject of 
theoretical$^{1-15}$ and 
experi\-mental$^{16-18}$ interest for many years.
Yet, there remain many unresolved questions: 1) Is there a difference
in the behaviors of the RFIM on varying the distribution of random
fields from, say, Gaussian to bimodal? 2) Does the transition in three
dimensions (3$D$) remain continuous down to zero 
temperature ($T$=0)? 3)Are there variants
of the RFIM that are in the same universality class?

In this note, we summarize the results of a scaling analysis of the RFIM
at $T$=0. We have studied three versions of the RFIM in 3$D$
and the bimodal and Gaussian RFIM in 4$D$. System sizes up to
$16 \times 16 \times 16$ and $10 \times 10 \times 10 \times 10$ have
been considered with the averaging carried out
over upto 10000 realizations (samples) of the random-field distribution.
For each sample, we obtained the ground state exactly. 
We find that all three variants of the RFIM
are probably in the same universality class in 3$D$ and estimate two of the 
exponents. In 4$D$, our results suggest that the Gaussian model has a
continuous transition whereas the bimodal model has a
first order transition at zero temperature. A novel finite-size scaling 
analysis is presented for this latter case. We discuss the physical
ramifications of our results at the end of this paper.

The RFIM is described by the Hamiltonian
\begin{equation}
{\cal{H}}_{RFIM}=-J \sum_{<ij>} S_i S_j - \sum_{i} h_i S_i \;\;,
\end{equation}
where $S_i=\pm 1$ and for the bimodal model $h_i=\pm h_r$ randomly
whereas the Gaussian model is characterized by a Gaussian distribution 
of $h_i$ with a width that we will call $h_r$. A variant of the RFIM that we
have studied in 3$D$ considers an Ising ferromagnet
in which a fraction of spins, $p/2$, is frozen to be $S_i=+1$
(in a quenched random manner) and an equal fraction ($p/2$) of spins
is frozen to be $S_i=-1$. We will denote this model as the random-pinned
(RP) model. All three models are identical and correspond to the Ising
ferromagnet when $h_r$ and $p$ are set equal to zero. Thus at $T$=0,
for $D$=3 and 4, the system is in a broken symmetry
state and has a magnetization equal to, say, +1. Our studies are
restricted to $T$=0, at which the magnetization goes to zero
at a threshold value of $h_r$ or $p$. Our focus is on studying this
transition by determining the ground states of the models exactly
for various values of the parameters. 

We note that the RP model
is a cousin of the spin one Ising model introduced by
Pereyra et al.$^{14}$
for a system of
N$_2$ molecules within a molecular crystal of CO molecules.
It is also related to the model of a dilute antiferromagnet
in a uniform field$^{16}$ 
with
$S_i=\pm 1$, but with a fixed fraction of the sites having no spin
(due to the random, quenched dilution). 

The three models that we
have studied are convenient in the sense that just one parameter needs to
be tuned at $T$=0 in order to reach the phase transition point.
This is in contrast to the recently introduced asymmetric generalization
of the random-field model for the description of the
liquid-vapor phase transition in porous media.$^{13}$ In that model, a fraction
$p$ of the sites have a field $+h'$, whereas the rest of the sites
have a field $-h''$, with the values of both $h'$ and $h''$ needing
fine tuning for a given $p\ne \frac{1}{2}$.
Our earlier studies have shown that a value of $p$ away from 
$\frac{1}{2}$ or correlations in the locations of the random field
lead to a tendency for the critical point at zero temperature
to be supplanted by a triple point at which three first order lines
intersect. The location of a zero-temperature critical point
(fixed point, in general) plays a key role in a scaling
description$^8$ of the random-field phase transition since it suggests
that the important competition is between the ordering tendency of the
exchange and the disordering tendency of the random field
with the role of temperature being secondary. The zero-temperature
fixed point scenario leads to measurable predictions$^8$ of a
violation of conventional hyperscaling and sluggish activated
dynamics. It is important to note, however, that the
absence of a zero-temperature critical point does not automatically
exclude the zero-temperature fixed point off in some point in
interaction space.

Within this context, it is interesting to note that within mean-field
theory,$^3$ the bimodal model and Gaussian model of the RFIM  behave
distinctly with the former having no $T=0$ critical point
unlike the latter. Our own studies of a bimodal model
on a Bethe lattice$^{13}$ show that on
varying the coordination number of the lattice,
both types of behavior are observed.

The simplest expectation in dimensions below the upper critical
dimension (ucd)  is that the bimodal model and the Gaussian model,$^{19}$
for which the ucd is 6, are both
in the same universality class and have a zero-temperature critical
point. Our results suggest that while this is possibly true in 3$D$,
differences arise even in 4$D$. We have used a polynomial-time
flow algorithm$^{20}$ first implemented in the study of the RFIM
by Ogielski$^{10}$ to determine the exact ground state.

Our analysis is based on studying the scaling behavior of the
Binder$^{21}$ parameter
$g=\frac{1}{2}[3\;-\; \frac{<m^4>}{<m^2>^2}] $,
where $m$ is the magnetization per site of a single realization and $<...>$ 
denotes a configurational averaging:
\begin{equation}
g_L(x)\;=\; g(L^{1/\nu}t) \;\;,
\end{equation}
with $t=x-x_c$,
where $x$ is $h_r$ for the bimodal and Gaussian models 
and $p$ for the RP model.
The transition value $x_c$ is first determined by
requiring that $g_L(x_c)$ is a constant independent of $L$.
We then collapse the $g_L(x)$ curves using the correlation length
exponent $\nu$ as an adjustable parameter. In order to determine the 
order-parameter exponent $\beta$, we use
\begin{equation}
<m^2>\;=\;L^{-2\beta /\nu}  {\psi}(L^{1/\nu}t)\;\;,
\end{equation}
with $\beta$ as an adjustable parameter.

For the smaller sizes considered we studied 10000 independent samples,
while for the larger systems we averaged over a few 1000 
independent realisations.
We note that our method does not involve problems of equilibration
encountered in Monte Carlo simulations, 
that the number of samples we have 
considered easily exceeds the usual numbers that are studied in
simulations and 
that the zero-temperature analysis accesses the random-field 
transition directly.

Typical scaling plots are shown in Figures 1-3.
The scaling relations are expected to hold for $L \rightarrow \infty$,
$t \rightarrow 0$, and $<m^2> \rightarrow 0$. However,
for the RFIM, $\beta$ is close to zero so that for the sizes
that we are able to study, $<m^2>$ is rather large. This is the
principal reason for the scaling not being better than it is.
The same fact precludes us from determining $\eta$. A measurement
of $\eta$ entails, e.g., a scaling collapse of $<m^2>$ 
at $t$=0 in the presence of 
a uniform field $H$ of varying strength. However, the large value
of $<m^2>$ at $t=H=0$ makes such a calculation inherently imprecise. 

While moderately good scaling is obtained for $g$, we have found that the
scaling of $<m^2>$ with predetermined values of $x_c$ and $\nu$
does not work well in both the $x < x_c$ and $x > x_c$ regimes
simultaneously. The best scaling plots are obtained on ignoring the 
$x < x_c$ region (since that is where $|m|$ is very large)
and choosing the optimal value of $\beta$ that collapses the
largest size systems best for $x > x_c$. Our results are summarized in
Table I. 

For $D=3$ the results (i.e.\ the values of $g_L(x_c)$ and the exponents $\beta$, 
$\nu$) are consistent with all models being in the same universality class. 
The extreme closeness of $g_c$ to unity implies that the distribution 
of $|m|$ at criticality is sharply peaked at a non-zero value. This is 
confirmed by inspecting this distribution directly (Figure 4). However, 
despite the smallness of $\beta$, there is no evidence that the transition is 
first order for $D=3$: the weight in $P(|m|)$ is concentrated around values 
of $|m|$ close to unity, without the additional peak at $|m|=0$ which would 
be expected if ferromagnetic and paramagnetic phases were to coexist 
at $h_r=h_c$. 

In $D=4$, the situation is very different.  The values of the exponent $\nu$ 
seem to be different for Gaussian and bimodal models; the exponent $\beta$ is 
demonstrably non-zero for the Gaussian model. The values of $g_c$ 
are different for the two field distributions
and an inspection of the whole 
distribution $P(|m|)$ (Figure 4) shows clearly that the Gaussian and bimodal 
models behave distinctly. 
While $P(|m|)$ for the Gaussian model has a single peak at $|m|>0$,
characteristic of a continuous transition, the bimodal model exhibits a 
{\em double peaked} distribution, with peaks at $|m|=1$ 
(not shown - see the caption to Figure 4) and $|m|=0$, suggestive 
of a discontinuous jump in $|m|$ at the transition (as predicted by mean
field theory$^3$).  

Finite-size scaling at a first-order transition requires the introduction 
of certain exponents analogous to critical exponents of continuous 
transitions.$^{22}$ \ We discuss briefly how this scaling analysis 
goes for a first-order random-field transition at $T=0$. For this 
transition, of course, $t = h_r - h_c$ plays the role of the `thermal' 
variable, while the ground-state energy $\langle E \rangle$ and its 
derivatives $d\langle E\rangle/dt$ and $d^2\langle E \rangle/dt^2$ play 
the roles of the free energy, the entropy and the specific heat respectively.  
Therefore, we anticipate that $d\langle E \rangle /dt$ as well as the 
magnetization $\langle m \rangle = - d\langle E \rangle /dH$ will be 
discontinuous at $h_c$. 
Now it is easily shown that $d\langle E \rangle /d(\ln h_r) = - \sum_ih_i S_i$. 
If one plots the distribution over samples of this quantity for $h_r=h_c$, 
one again finds a double-peaked distribution for the 4D bimodal model (but 
single peaked for the other models). A peak at the origin corresponds to 
the samples with most of the spins aligned, and  there is a broad second 
peak corresponding to the non-zero weight at small $|m|$ in Figure 4. 
We infer that $d\langle E \rangle /dt$ is discontinuous at $t=0$ as 
anticipated.  

The finite-size scaling analysis starts from the scaling form for the 
singular part of the configuration-averaged energy density, 
$\langle E(t,H) \rangle = L^{y-D} f(tL^{1/\nu_t},HL^{1/\nu_H})$, 
where $H$ is a uniform magnetic field, and $y$ is the scaling dimension of 
the Hamiltonian at the transition.$^8$ \ Discontinuities in the derivatives 
with respect to $t$ and $H$ in the limit $L \to \infty$ imply  
$1/\nu_t = 1/\nu_H = D-y$. This generalizes the result 
$1/\nu_t = 1/\nu_H = D$ of conventional first-order transitions at 
$T_c>0$.$^{22}$ \ At the transition, the (connected) susceptibility 
$\chi_{con} = \partial \langle m \rangle /\partial H \sim L^{D-y} 
= L^{2-\eta}$, giving $\eta = 2-D+y$. The exponent $\bar{\eta}$ is 
defined through the `disconnected' susceptibility 
$\chi_{dis} = L^D \langle m^2 \rangle \sim L^{4-\bar{\eta}}$, giving 
$\bar{\eta} = 4-D$, because $\langle m^2 \rangle$ jumps at the transition.  
Finally one can readily derive a Schwartz-Soffer inequality 
$\bar{\eta} \le 2\eta$, following the method used for continuous 
transitions.$^{23}$ \ Together with the scaling relations derived above, 
this implies $y \ge D/2$ and $\nu_t =\nu_H \ge 2/D$, 
reminiscent of a general inequality for random systems.$^{24}$
For $D=4$ we obtain $\nu_t \ge 1/2$, consistent with our numerical result 
$\nu=0.6 \pm 0.1$. If, as has been suggested,$^{25}$ the Schwartz-Soffer 
inequality is saturated this becomes $\nu_t=1/2$, still consistent with the 
numerical result. 

Our results have implications for experimental realizations of the
RFIM. It is likely that for situations such as the liquid vapor
transition in aerogel,$^{18}$ the correlations in the strands as well
as the large porosity cause the $T$=0 critical point to be
supplanted by a first order transition.$^{13}$ The key question then
is what the exponents are for the continuous transition at
non-zero temperature when the random-field strength is weak.
The simplest scenario is that the governing fixed point
is still at $T$=0 and is characterized by conventional random-field 
exponents but this has not been explicitly demonstrated yet.
A more intriguing scenario would correspond to an entirely new
universality class. In the former case, the as yet unexplained results
of Wong and Chan$^{18}$ could perhaps be attributed to the experiments
not being carried out sufficiently close to the critical point to see
the crossover from bulk-like behavior to the ``true" random-field 
behavior.

We are indebted to Moses Chan for stimulating discussions.
This work was supported by
grants from EPSRC, KBN (grant number 2P302-127, Poland), 
INFN (Italy), NASA, NATO, 
a NSF MRG grant, and the Petroleum
Research Fund administered by the American 
Chemical Society. MRS acknowledges the E. U., contract ERB CHB
GTC 940636, for financial support.

\newpage

\begin{center}
REFERENCES
\end{center}
\begin{description}

\item {1. For reviews, see T. Natterman and P. Rujan, Int.\ J. Mod.\ Phys.\
B{\bf 3}, 1597 (1989); D. S. Fisher, G. M. Grinstein, and A. Khurana,
Phys.\ Today {\bf 41}, (12), 58 (1988).}

\item {2. Y. Imry and S. K. Ma, Phys.\ Rev.\ Lett.\ {\bf 35}, 1399 (1975).}

\item {3. T. Schneider and E. Pytte, Phys.\ Rev.\ B {\bf 15}, 1519 (1977).
For a general analysis see A. Aharony, Phys.\ Rev.\ B {\bf 18}, 3318 (1978).}

\item {4. D. Andelman, Phys.\ Rev.\ B {\bf 27}, 3079 (1983);
M. Kaufman, P. E. Klunzinger, and A. Khurana, Phys.\ Rev.\ B {\bf 34},
4766 (1986); T. Horiguchi, J. Math.\ Phys.\ {\bf 20}, 1774 (1979).}

\item {5. R. Bruinsma, Phys.\ Rev.\ B {\bf 30}, 289 (1984).}

\item {6. A. Houghton, A. Khurana, and F. J. Seco, Phys.\ Rev.\ Lett.\ 
{\bf 55}, 856 (1985).}

\item {7. A. P. Young and M. Nauenberg, Phys.\ Rev.\ Lett.\ {\bf 54}, 2429
(1985).}

\item {8. A. J. Bray and M. A. Moore, J. Phys.\ C {\bf 18}, L927 (1985);
J. Villain, J. Physique {\bf 46}, 1843 (1985); D. S. Fisher,
Phys.\ Rev.\ Lett.\ {\bf 56}, 416 (1986).}

\item {9. A. T. Ogielski and D. A. Huse, Phys.\ Rev.\ Lett.\ {\bf 56},
1298 (1986).}

\item {10. A. Ogielski, Phys.\ Rev.\ Lett.\ {\bf 57}, 1251 (1986); 
we are grateful to Andrew Ogielski for providing us with a copy of ref. 20.
Our calculations are similar in spirit to Ogielski's -- the scaling
analysis is somewhat different. Our results are consistent
with his for the 3$D$ Gaussian model.}

\item {11. G. S. Grest, C. M. Soukoulis, and K. Levin, Phys.\ Rev.\ B 
{\bf 33}, 7659 (1986).}

\item {12. S. R. McKay and A. N. Berker, J. Appl.\ Phys.\ {\bf 64}, 5785
(1988); S. R. McKay and A. N. Berker, in {\it New Trends in Magnetism}, ed.
M. D. Coutinho-Filho and S. M. Rezende, World Scientific, Singapore (1989);
M. S. Cao and J.Machta, Phys.\ Rev.\ B {\bf 48}, 3177 (1993); A. Falicov, 
A. N. Berker and S. R. McKay, Phys.\ Rev.\ B {\bf 51}, 8266 (1995).}

\item {13. A. Maritan, M. R. Swift, M. Cieplak, M. H. Chan, M. W. Cole, 
and J. R. Banavar, Phys.\ Rev.\ Lett.\ {\bf 67}, 1821 (1991);
M. R. Swift, A. Maritan, M. Cieplak, and J. R. Banavar, J. Phys.\ A 
{\bf 27}, 1525 (1994).}

\item {14. V. Pereyra, P. Nielaba, and K. Binder, Z.\ Phys. \ {\bf B97},
179 (1995).}

\item {15. H. Rieger and A. P. Young, J. Phys.\ A {\bf 26}, 5279 (1993);
H. Rieger, Phys. \ Rev. \ {\bf B52}, 6659 (1995); 
M. E. J. Newman and G. T. Barkema, Phys. \ Rev. \ {\bf B53} 393 (1996).}

\item {16. S. Fishman and A. Aharony, J. Phys.\ C {\bf 12}, L729 (1979);
P. Z. Wong, S. von Molnar, and P. Dimon, J. Appl.\ Phys.\ 
{\bf 53}, 7954 (1982); R. J. Birgeneau, R. A. Cowley, G. Shirane,
and H. Yoshizawa, Phys.\ Rev.\ Lett.\ {\bf 48}, 1050 (1982);
J. Cardy, Phys.\ Rev.\ B {\bf 29}, 505 (1984);
J. P. Hill, T. R. Thurston, R. W. Erwin, M. J. Ramsted, and R. J. Birgeneau,
Phys.\ Rev.\ Lett.\ {\bf 66}, 3281 (1991); P.-Z. Wong, Phys. Rev. Lett.
{\bf 77}, 2338 (1996); J. P. Hill, Q. Feng and R. J. Birgeneau, 
Phys. Rev. Lett. {\bf 77}, 2339 (1996);  P.-Z. Wong, Phys. Rev. Lett.
{\bf 77}, 2340 (1996); D. P. Belanger, W. Kleemann and F. C. Montenegro,
Phys. Rev. Lett. {\bf 77}, 2341 (1996); R. J. Birgeneau, Q. Feng, Q. J.
Harris, J. P. Hill and A. P. Ramirez, Phys. Rev. Lett. {\bf 77}, 2342 (1996).}

\item {17. P. G. de Gennes, J. Phys.\ Chem., {\bf 88}, 6469 (1984);
J. V. Maher, W. I. Goldburg, D. W. Pohl, and M. Lanz, Phys.\ Rev.\ Lett.\ 
{\bf 53}, 60 (1984); M. C. Goh, W. I. Goldburg, and C. M. Knobler, Phys.\ 
Rev.\ Lett.\ {\bf 58}, 1008 (1987); P. Wiltzius, S. B. Dierker, and 
B. S. Dennis, Phys.\ Rev.\ Lett.\ {\bf 62}, 804 (1989).}

\item {18. A. Wong and M. H. W. Chan, Phys.\ Rev.\ Lett.\ 
{\bf 65}, 2567 (1990).} 

\item {19. Y. Imry, S. - K. Ma, Phys. Rev. Lett. {\bf 35}, 1399 (1975);
A. Aharony, Y. Imry, and S. - K. Ma, Phys. Rev. Lett. {\bf 37}, 1364 (1976).}

\item {20. A. V. Goldberg, M. I. T. Report, November 1985.}

\item {21. K. Binder, in {\it Computational Methods in Field Theory},
ed. by H. Gausterer and C. B. Lang (Berlin, Springer 1992).}

\item {22. M. E. Fisher and A. N. Berker, Phys.\ Rev.\ B {\bf 26}, 2507 
(1982). }

\item {23. M. Schwartz and A. Soffer, Phys.\ Rev.\ Lett.\ {\bf 55}, 2499 
(1985). }

\item {24. J. T. Chayes, L. Chayes, D. S. Fisher and T. Spencer, 
Phys.\ Rev.\ Lett.\ {\bf 57}, 2999 (1986). }

\item {25. M. Gofman, J. Adler, A. Aharony, A. B. Harris, and M. Schwartz, 
Phys.\ Rev.\ Lett.\ {\bf 71}, 1569 (1993). }

\end{description}

\newpage

\begin{center}
{\underline{\LARGE TABLE I}}
\end{center}

\begin{center}
\begin{tabular}{|c|c|c|c|c|}
\hline
\\

MODEL   &    $ h_c$ or $p_c$ &     $  g_c $ & $   \beta$   &$  \nu$\\
\\
\hline
\\

Bimodal 3$D$ & $2.25\pm 0.02$ & $0.996\pm 0.002$
&  $0.025\pm 0.015$ & $1.2\pm 0.15$\\  \\

Gaussian 3$D$  & $2.33\pm 0.03$ & $0.997\pm 0.002$ & 
$0.031 \pm 0.015$ & $1.2\pm0.15$\\  \\

RP 3$D$& $ 0.19\pm 0.01$ & $0.997\pm 0.002$
&$ 0.025\pm 0.02$ & $1.2\pm 0.2$ \\  \\

Bimodal 4$D$ & $3.71\pm 0.05$ & $0.80\pm 0.05$
&$ 0 (< 0.01) $ & $0.6 \pm 0.1 $ \\  \\

Gaussian 4$D$ &$ 4.17\pm 0.05$ & $0.96\pm 0.05$ & $0.13\pm 0.02$
&$ 0.8 \pm  0.1 $ \\  \\
\hline
\end{tabular}
\end{center}

\vspace{1cm}

\noindent
Table Caption: Summary of the results obtained in this paper.

\newpage

\noindent
FIGURE CAPTIONS

\begin{description}

\item{1. The scaling plots for the $D$=3 Gaussian model
for the parameters shown in Table I. The values of $L$
studied are 6, 8, 12, and 16. The symbols with more sides
correspond to larger values of $L$.}

\item{2. Same as Figure 1 but for the 4$D$ Gaussian system.
Here $L$=4, 5, 6 and 10.}

\item{3. Same as Figure 2 but for the 4$D$ bimodal system
with $L$=5, 6, 8 and 10.}

\item{4. Probability distributions for the absolute value of magnetization
for the Gaussian and bimodal 3- and 4-$D$ systems
(for $L$=16 and 10 respectively) at the phase transition (Table I).
In the bimodal case, the bin 
with $|m|$=1 is not shown.
80\% of the samples in 3$D$ and 64\% in 4$D$ had $|m|$=1.
On increasing the system size, the peaks become more pronounced
while retaining the same structure$^{15}$.}

\end{description}

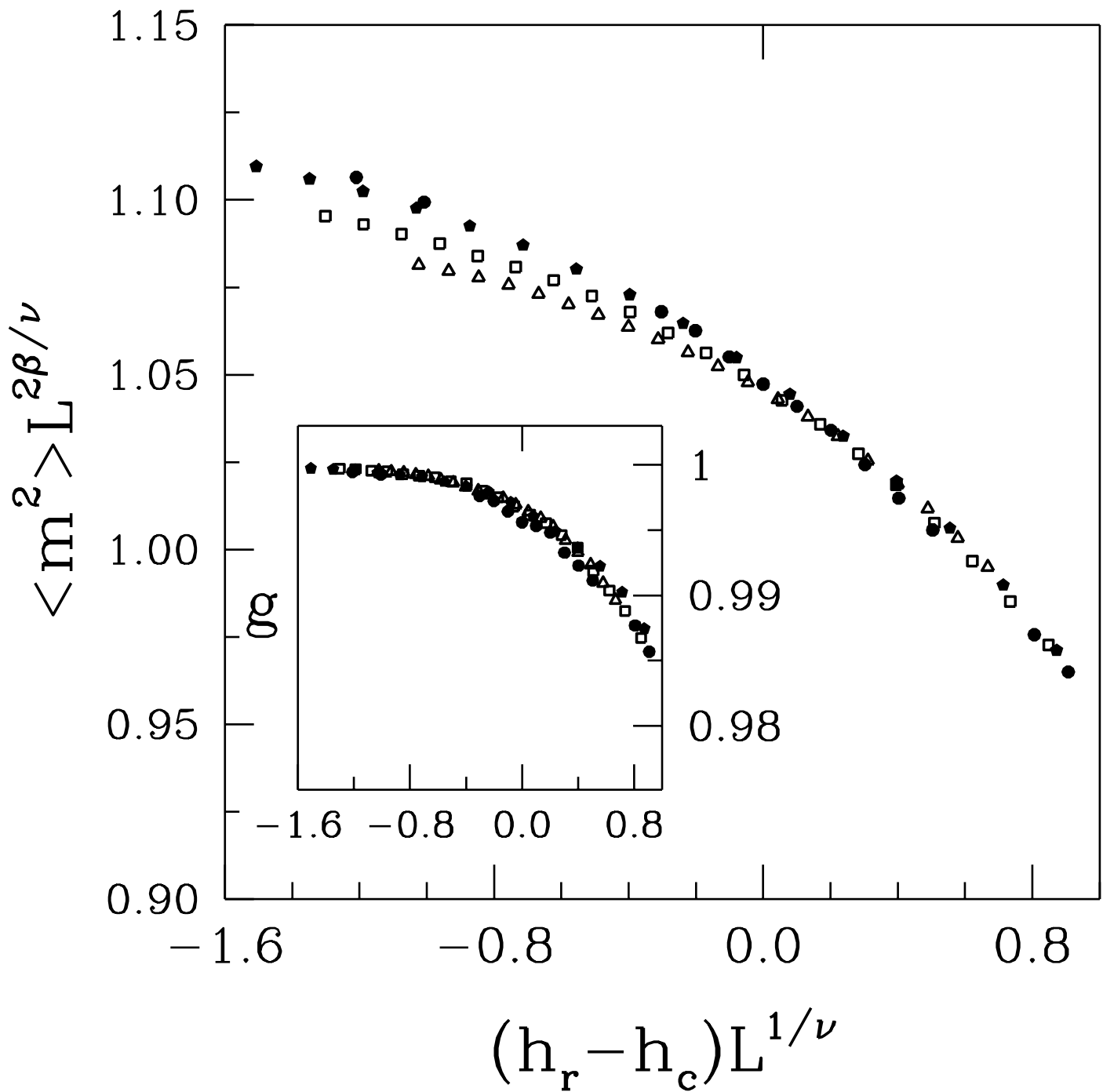
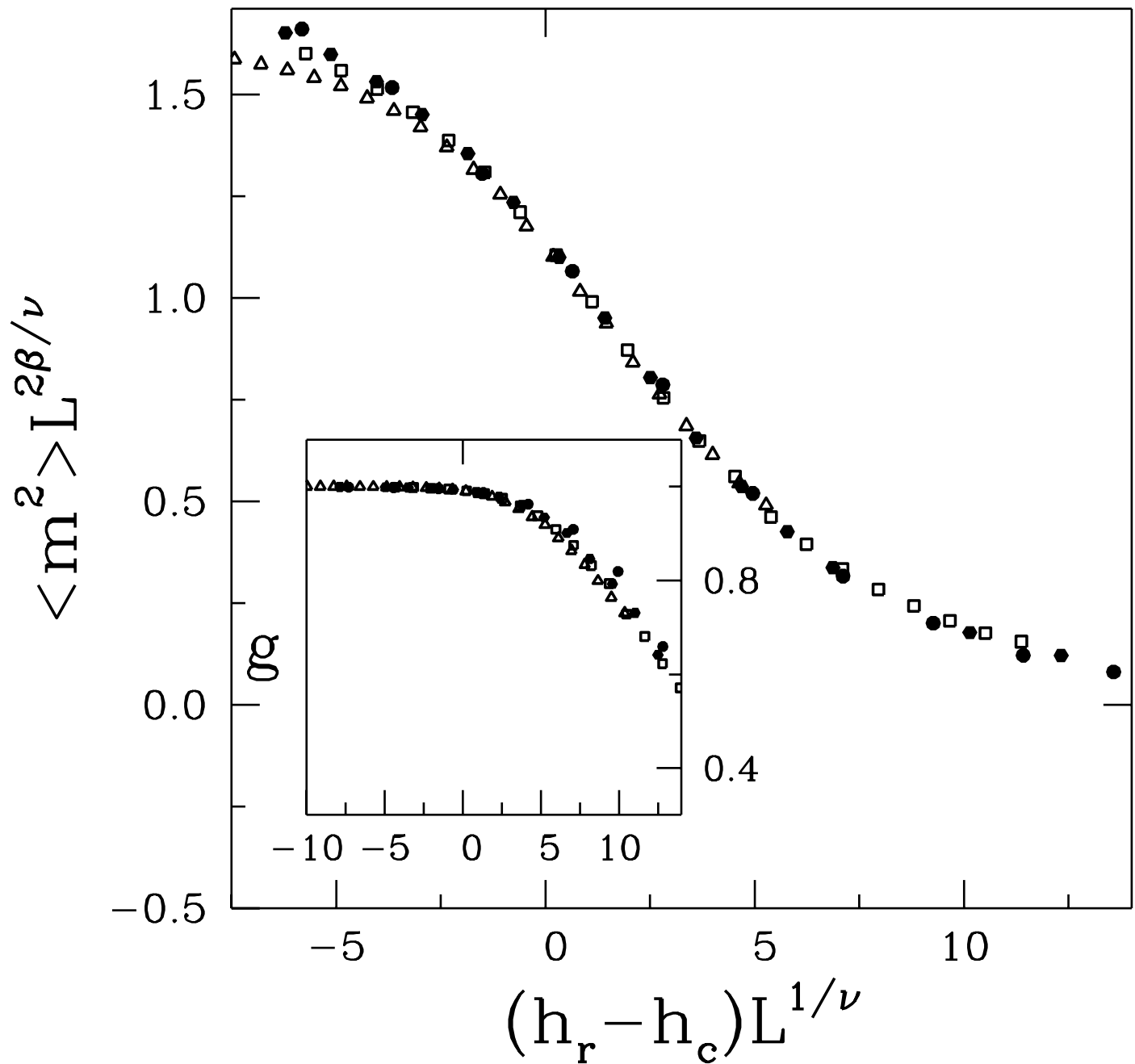
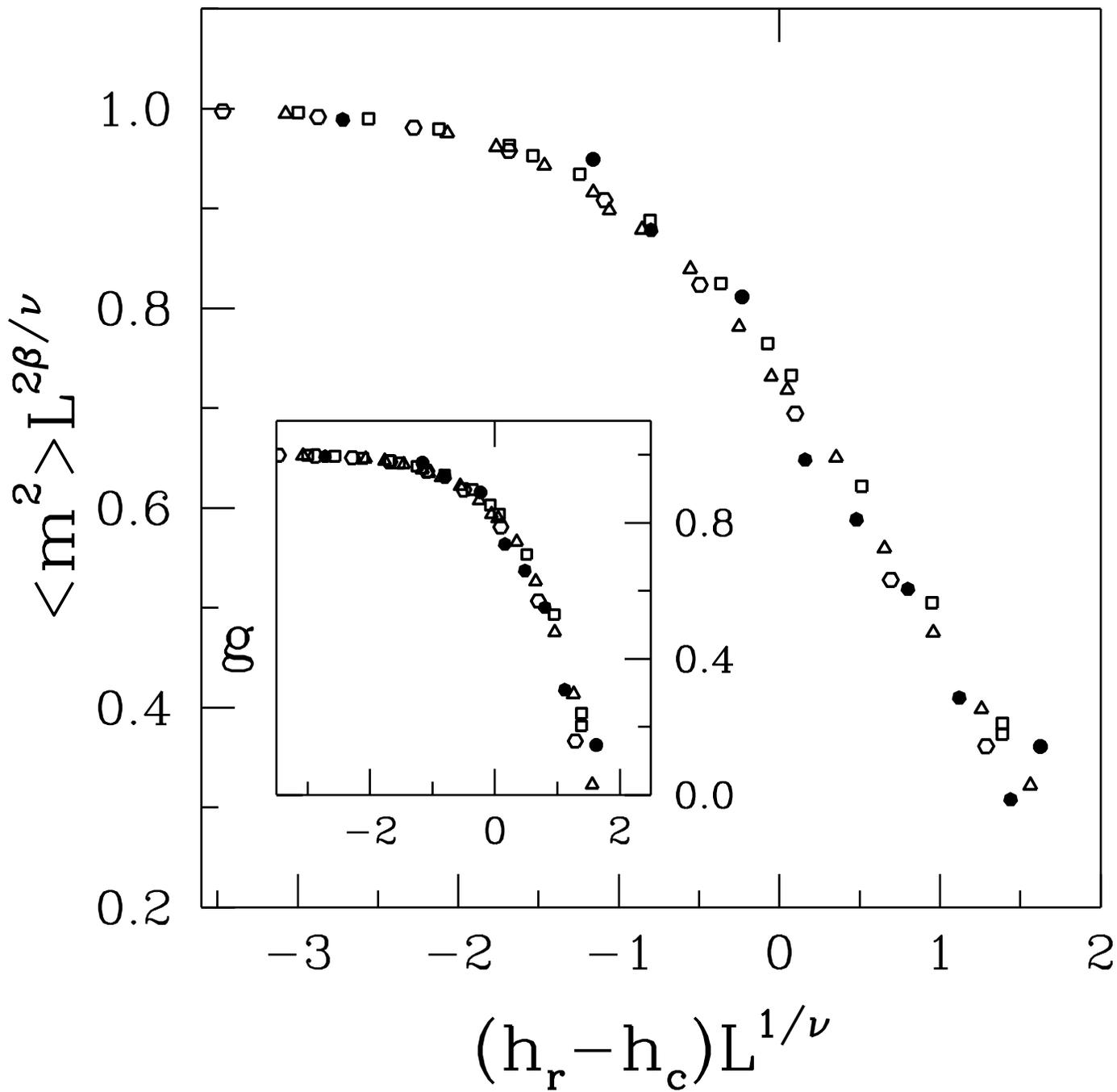
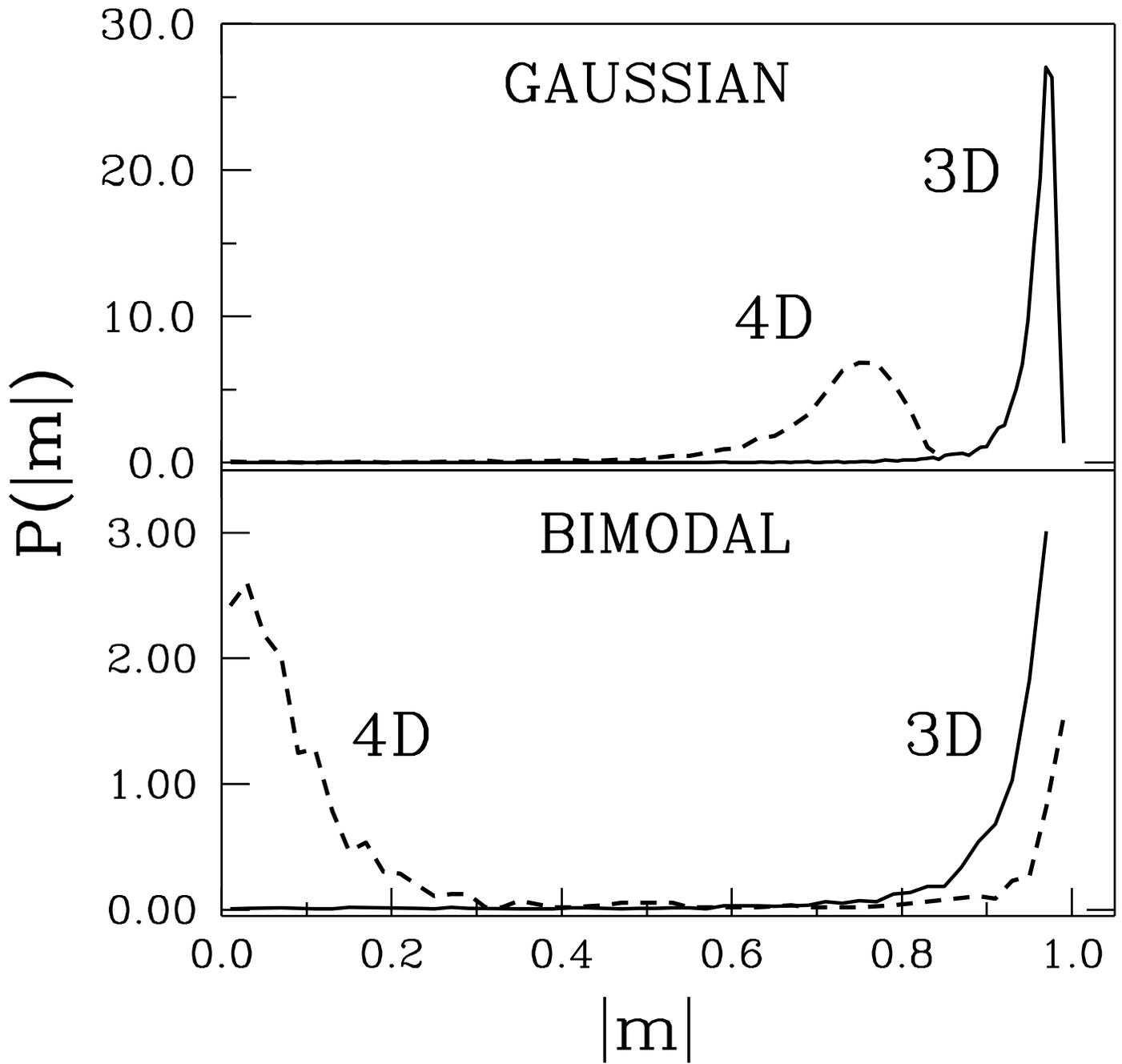
\end{document}